# Decoding Superconductivity in $La_3Ni_2O_{7-\delta}$ Thin Films via Ozone-Driven Structure and Oxidation Tuning

*Mathieu Flavenot, Hoshang Sahib, Jérôme Robert, Marc Lenertz, Gilles Versini, Laurent Schlur, Alexandre Gloter\*, Nathalie Viart* and *Daniele Preziosi\**

M. Flavenot, H. Sahib[⊥], J. Robert, M. Lenertz, G. Versini, L. Schlur, N. Viart and D. Preziosi
Université de Strasbourg, CNRS, IPCMS UMR 7504, F-67034 Strasbourg, France
Email Address: daniele.preziosi@ipcms.unistra.fr
A. Gloter
Laboratoire de Physique des Solides, CNRS, Universite Paris-Saclay, 91405 Orsay, France
Email Address: alexandre.gloter@universite-paris-saclay.fr
[⊥] Present Address: Department of Physics, College of Science, University of Halabja, Halabja, Iraq

The discovery of superconductivity in bulk Ruddlesden-Popper $La_3Ni_2O_7$(LNO327) under high hydrostatic pressure has redefined the recent experimental consensus that nickelate superconductivity is restricted to systems with a $3d^9$ electronic configuration and square-planar coordination. However, the structural and electronic prerequisites for stabilizing superconductivity, whether under pressure or at ambient conditions in the case of thin films, remain poorly understood, largely due to the metastable nature of the LNO327 phase. Here, we present a detailed structural study of epitaxial $La_3Ni_2O_{7-\delta}$ thin films by using scanning transmission electron microscopy (STEM) combined with electron energy loss spectroscopy (EELS). Grown via pulsed laser deposition onto $SrLaAlO_4$ substrates, those films exhibit distinct superconducting properties as a function of the different post-annealing conditions used. By correlating the rich landscape of stacking polymorphs with transport behavior, this work establishes a framework for understanding the metastable superconducting phase in bilayer nickelate thin films. Our findings underscore the critical role of homogeneity in oxygen stoichiometry, epitaxial strain and structural motif in stabilizing superconductivity, offering a clear pathway for designing ambient-pressure superconducting nickelates.

## 1 Introduction

The recent discovery of superconductivity below 80 K in bulk Ruddlesden-Popper (RP) $La_{n+1}Ni_nO_{3n+1}$ with $n = 2$ under high hydrostatic pressure (above 14 GPa) has reignited intense interest in nickel-based superconductors [1, 2, 3]. This breakthrough marks a new chapter in the so-called "Nickel Age", demonstrating superconductivity in a layered compound with a bilayer structure and a nominal $Ni^{2.5+}$ valence state ($3d^{7.5}$), clearly distinct from both cuprates and infinite-layer (IL) nickelates [4]. In contrast to IL nickelates, where Ni ions adopt a $3d^9$ configuration with holes primarily residing in the $Ni-3d_{x^2-y^2}$ orbitals [5], the bilayer $La_3Ni_2O_7$ (LNO327) exhibits a multi-orbital electronic structure. In particular, the $Ni-3d_{z^2}$ orbitals play a relevant role by enabling interlayer electronic hopping between adjacent $NiO_2$ planes inside the bilayer, which is believed fundamental to stabilize the superconducting state. Recent spectroscopic investigations combined with double-cluster calculations, suggest that LNO327 is best described by a predominant $Ni-3d^8$ electronic configuration accompanied by a significant density of ligand holes [6, 7]. This mixed-valence character originates from a relatively small charge-transfer energy ($< 2\,\text{eV}$), placing the LNO327 in an intermediate regime between Mott-Hubbard and charge-transfer physics within the Zaanen-Sawatzky-Allen classification [8].
At ambient conditions, LNO327 crystallizes in the orthorhombic *Amam* structure (a=5.39 Å, b=5.45 Å, c=20.54 Å), consisting of double perovskite $NiO_2$ layers separated by rock-salt LaO layers along the out-of-plane direction (*cf.* Figure 1a). This structure corresponds to a bilayer (BL) RP stacking sequence, often denoted as LNO-2222, in which $NiO_6$ octahedra are corner-sharing within the planes and connected along the *c*-axis, while the LaO layers act as spacers. The octahedra are tilted, leading to Ni-O-Ni bond angle values of roughly 168°. Under hydrostatic pressure, a progressive straightening of the apical Ni-O-Ni bond angle toward 180° is observed, signaling a structural transition that has been closely linked to the emergence of superconductivity [9, 10]. While early high-pressure x-ray diffraction (XRD) studies suggested a tetragonal *I4/mmm* symmetry for the superconducting phase [11], more recent experiments performed under improved hydrostatic conditions indicate that an orthorhombic $Fmmm$ structure provides a more accurate description of the structural data [9]. Notably, bulk samples stabilized in the tetragonal phase at ambient pressure do not exhibit superconductivity even under pressures up to 68 GPa, highlighting a strong correlation between structural distortions and superconductivity. However, it remains an open question whether the orthorhombic distortion itself is essential for pairing, or whether additional mechanisms cooperate with the structural instability. Those uncertainties are further exacerbated by a critical challenge: the metastable nature of the LNO327 phase, which promotes the formation of alternative stacking sequences and structural intergrowths [12, 13]. In particular, intergrowths with other RP phases such as $La_2NiO_4$ ($n=1$) and $La_4Ni_3O_{10}$ ($n=3$) are frequently observed [14]. Of particular interest is a polymorphic stacking sequence consisting of alternating $La_2NiO_4$ monolayers (ML) and $La_4Ni_3O_{10}$ trilayers (TL), commonly referred to as LNO-1313, which preserves the overall $La_3Ni_2O_7$ stoichiometry. Remarkably, this polymorph has also been reported to exhibit superconductivity under high pressure [15, 16], fueling an ongoing debate regarding the structural phase required for superconductivity. Chemical substitution on the La site with smaller rare-earth elements (*e.g.*, Pr, Sm), was found to stabilize the LNO-2222 phase via chemical pressure, improving superconducting properties, although under relatively high pressure conditions [14, 17].

More recently, superconductivity at ambient pressure with a decreased onset $T_c$ around 42 K has been reported also in epitaxial LNO327 ultrathin films under in-plane compressive strain and following ozone ($O_3$) annealing [18]. However, the superconducting phase in thin films remains highly unstable and strongly dependent on growth and post-treatment conditions. This instability highlights the extreme sensitivity of LNO327 to oxygen stoichiometry, particularly at the apical oxygen sites that control interlayer coupling and Ni-3$d$ orbital polarization. In the absence of hydrostatic pressure, the superconducting phase results from a delicate interplay between epitaxial strain and oxygen content, making it intrinsically metastable and prone to degradation (oxygen vacancies). In this context, a direct and unambiguous structural characterization of superconducting LNO327 thin films with different critical temperatures ($T_c$) is needed.

In this work, we present a comprehensive structural study via scanning transmission electron microscopy (STEM) with electron energy loss spectroscopy (EELS) of a series of compressively strained epitaxial $La_3Ni_2O_{7-\delta}$ thin films exhibiting distinct superconducting properties and onset $T_c$ values at the time of transport characterization. The thin films were grown onto (001)-oriented $SrLaAlO_4$ (SLAO) substrates by pulsed laser deposition (PLD) assisted by reflective high energy electron diffraction (RHEED) and the film thickness was kept below 6 nm to allow the compressive strain to be retained up to the SLAO capping layer (please refer to the Experimental Section for further details). High-resolution STEM measurements acquired with a high-angle annular dark-field (HAADF) imaging method, allowed us to identify the stacking sequences and structural polymorphs associated with the different defect landscapes in each thin film. While the superconducting state may not be preserved during or after STEM specimen preparation, we believe that the structural information obtained here provides a reliable description of the phases and stacking motifs stabilized by the growth and post-treatment conditions. By comparing films with distinct transport behaviors, our results establish a structural reference framework for LNO327 thin films and highlight the range of structural configurations accessible in this system (LNO-2222 vs LNO-1313), offering new insight into the structure that possibly stabilize superconductivity at ambient pressure and guiding future efforts to engineer more stable superconducting nickelates.

## 2 Results and Discussion

The growth of LNO327 thin films under relatively large compressive strain is a delicate process as the material tends to loose oxygen under the extreme temperature and atmosphere conditions typically required for the growth. The resulting oxygen vacancies, not only make the films sensitive to red-ox reactions during synthesis, but provide also a kinetic pathway for stabilizing thermodynamically more stable, yet undesired, phases. Therefore, while distinguishing between the $Amam$ and $Fmmm$ space groups in thin films remains experimentally challenging, the successful stabilization of the orthorhombic crystal structure within those limitations represents *per se* a critical step to achieve a superconducting ground state.

## 2.1    Growth and Macroscopic Structural Quality

We begin by presenting in **Figure 1** the structural and morphological properties of a representative SLAO-capped 6 nm thick La$_3$Ni$_2$O$_{7-\delta}$ epitaxial film, after a thorough optimization of the growth parameters (details not shown). The x-ray diffraction (XRD) $\theta - 2\theta$ scan displays sharp (00$l$) diffraction peaks without any detectable impurity or secondary phases, indicating the high crystallinity and phase purity of the film, which can be unambiguously indexed as orthorhombic LNO327. From the peak positions, a $c$-axis lattice parameter of 20.8(2) Å can be calculated, larger than the bulk value of 20.54 Å, and therefore consistent with an overall in-plane compressive strain state. Within experimental uncertainty, this value is highly reproducible across a series of nominally identical samples which are grown under the same conditions. Reciprocal space mapping (RSM) performed around the (1 0 11) and (1 1 17) reflections of the substrate and LNO327 film, respectively, further confirm the high degree of epitaxial strain (Figure 1c) and reveals an in-plane lattice parameter of 3.76 Å. The structural quality of the surface and near-surface region is further supported by RHEED patterns acquired after the LNO327 film deposition (Figure 1d). The streaky two-dimensional diffraction features, together with the presence of well-defined Kikuchi lines, indicate a smooth surface morphology and are characteristic of a layer-by-layer growth mode of the LNO327 thin film. Atomic force microscopy (AFM) measurements performed on the complete capped sample (Figure 1e) reveal a flat surface with an overall root-mean-square roughness of approximately 0.6 nm confirming the high-quality of the subsequent SLAO capping layer. Taken together, those results demonstrate that the La$_3$Ni$_2$O$_{7-\delta}$ films are coherently strained and exhibit high crystallographic and morphological quality.

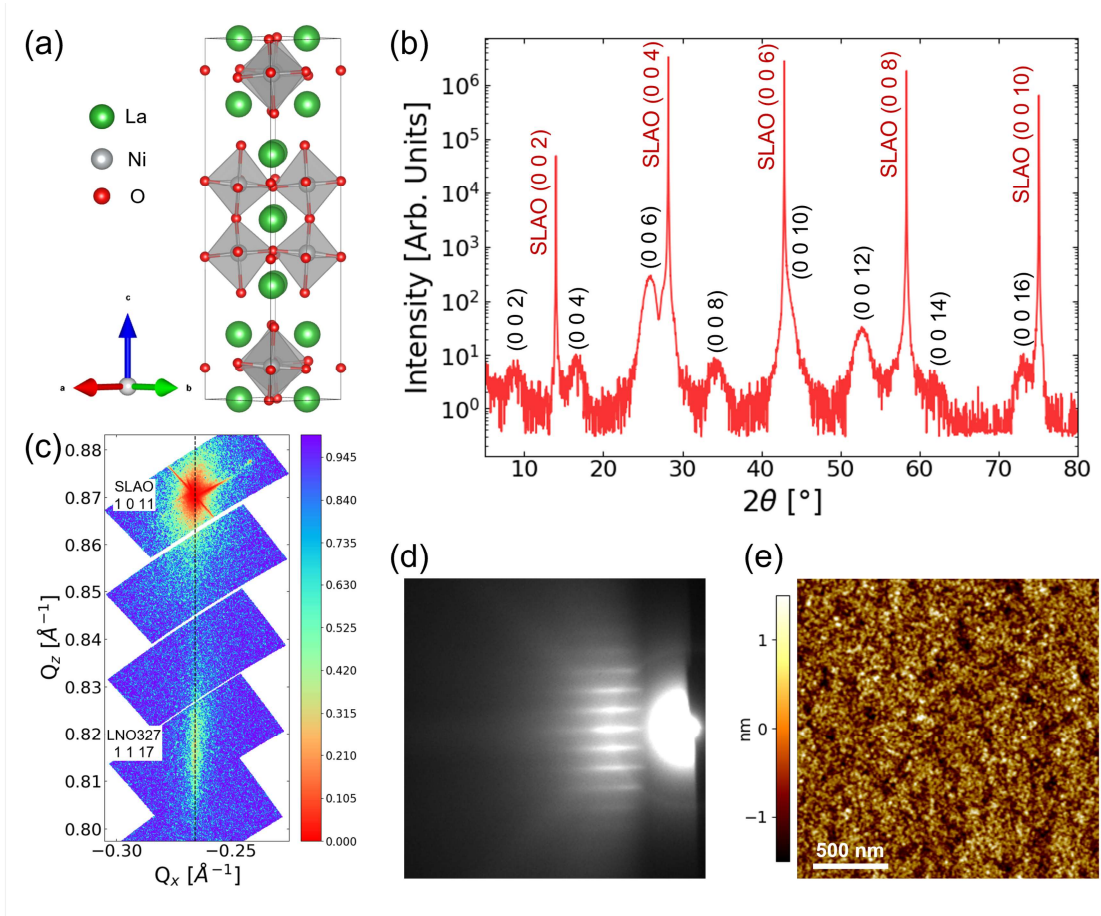

Figure 1: (a) *Amam* orthorhombic crystal structure of La$_3$Ni$_2$O$_7$. (b) X-ray diffraction pattern of a representative La$_3$Ni$_2$O$_7$ thin films onto a (001)-oriented SrLaAlO$_4$ substrate. The characteristic peaks of the La$_3$Ni$_2$O$_7$ film and the SrLaAlO$_4$ substrate are labelled as (00l). (c) Reciprocal space mapping around the (1 0 11) and (1 1 17) peaks of the substrate and film, respectively. (d) Reflective high energy electron diffraction after the growth of the La$_3$Ni$_2$O$_7$ layer. (e) Atomic force microscopy image acquired after deposition of the SLAO capping layer.

## 2.2 Transport Properties after Ozone Annealing

All as-grown samples display an insulating transport behavior, mostly due to their poor degree of oxidation (please refer to the SI). As a result, a post-growth $O_3$ annealing step is necessary not only to achieve metallic conductivity, but also to induce a superconducting transition, as previously demonstrated in the literature [18, 19, 20, 21]. This clearly evidences the crucial role of oxygen stoichiometry in governing the transport properties of $La_3Ni_2O_{7-\delta}$ thin films. In this work, three nominally identical films were grown and subsequently subjected to ex situ post-growth $O_3$ annealing under different conditions, as summarized in Table 1 and hereafter referred to as S1, S2 and S3 (please refer to the Experimental Section for further details about the annealing procedure). **Figure 2** presents the corresponding transport, magnetotransport and XRD measurements following $O_3$ annealing. After the $O_3$ annealing, all samples exhibit metallic behavior at high temperatures, characterized by a monotonic decrease of resistance with decreasing temperature. Upon further cooling, a clear deviation from the normal-state trend is observed, marked by a broad resistivity downturn indicative of the onset of superconducting correlations (Figure 2a). The magnitude and temperature dependence of this low-temperature resistivity downturn are strongly dependent on the dynamics of the oxygen incorporation during annealing which we could control by opportunely varying the $O_3$-concentration, annealing temperature and process duration. In particular, sample S3, annealed at a relatively low temperature for a shorter time with respect to the other samples, displays a significantly higher normal-state resistivity, together with a less pronounced low-temperature downturn and clear indications of transport non-homogeneity (please refer to the SI). As a result of progressively modified oxidizing conditions, samples S1 and S2 exhibit a lower resistivity and a more pronounced superconducting downturn together with an increased spatial homogeneity. In particular, sample S1, which was treated with slightly (nominally) lower $O_3$ concentration and temperature than S2, nearly achieves a zero-resistance state. The superconducting downturn temperatures (*i.e.* onset $T_c$, further details are provided in the SI) for the S1 and S2 samples are summarized in Table 1. The superconducting transition of sample S1 extends over a relatively broad temperature range, suggesting the presence of strong superconducting fluctuations and/or spatial non-homogeneity at the microscopic level, mostly to be reconducted to a not optimal $O_2$ incorporation. Although a zero-resistance state is not achieved down to the lowest reachable temperature of our transport system (2 K), the sharp resistance reduction resistance at low temperature, together with its sensitivity to magnetic field, supports the emergence of a superconducting state. The superconducting nature of the transition is, indeed, further corroborated by magnetotransport measurements, which reveal a systematic suppression and decrease of the onset $T_c$ values upon application of a magnetic field perpendicularly to the sample surface (*cf.* Figure 2b,c), as already shown in literature for similar thin films [18]. The $T_c$ follows a linear decrease with the magnetic field consistent with the linearized Ginzburg-Landau model for a magnetic field applied perpendicularly to the sample surface and in agreement with literature reports on superconducting nickelates [4, 18, 19, 20, 21, 22]. Moreover the extracted upper critical fields of 87 T and 25 T for S1 and S2 point out a more robust superconductivity for S1 (please refer to the SI). Furthermore, consistently with previous reports on undoped LNO327 thin films [18, 23], our Hall effect measurements confirm that holes are the majority charge carriers, with their concentration increasing as the temperature approaches the superconducting onset $T_c$ (see SI). Figure 2d shows the XRD patterns of the samples after the $O_3$ annealing treatment. Sample S1 retains a well-defined orthorhombic LNO327 crystal structure, as evidenced by the sharp and clearly resolved diffraction peaks characteristic of the 2222 stacking sequence, without any significant change in its c axis parameter (20.8(1) Å). On the contrary, samples S2 and S3 exhibit a degradation of their diffraction patterns, manifested by an overall reduction of the peak intensities and a decrease of their c axis parameter (20.7(2) Å and 20.6(1) Å, respectively), indicating a strong sensitivity of the 2222 polymorph to oxygen dynamics exchange. This effect is particularly pronounced for sample S3, where several weaker reflections, such as the (002), (004), (008) and (0014), are no longer detectable within the sensitivity of the measurement. The observed changes in the diffraction features point to possible variations in the local stacking sequence, defect density and eventually oxygen stoichiometry as well. The degraded XRD patterns observed for samples S2 and S3, in contrast to the largely unaltered S1, suggest that the different $O_3$ annealing conditions induce subtle structural modifications when the annealing conditions are not optimal. Since such nanoscale structural inhomogeneities can strongly influence the transport properties, a more direct probe of the local microstructure is required. Therefore, we resorted to high-resolution STEM measurements supported by spatially-resolved EELS maps to investigate the atomic-scale structure of each films with the main goal to establish a direct link between the transport behavior and the microstructural characteristics of our samples.

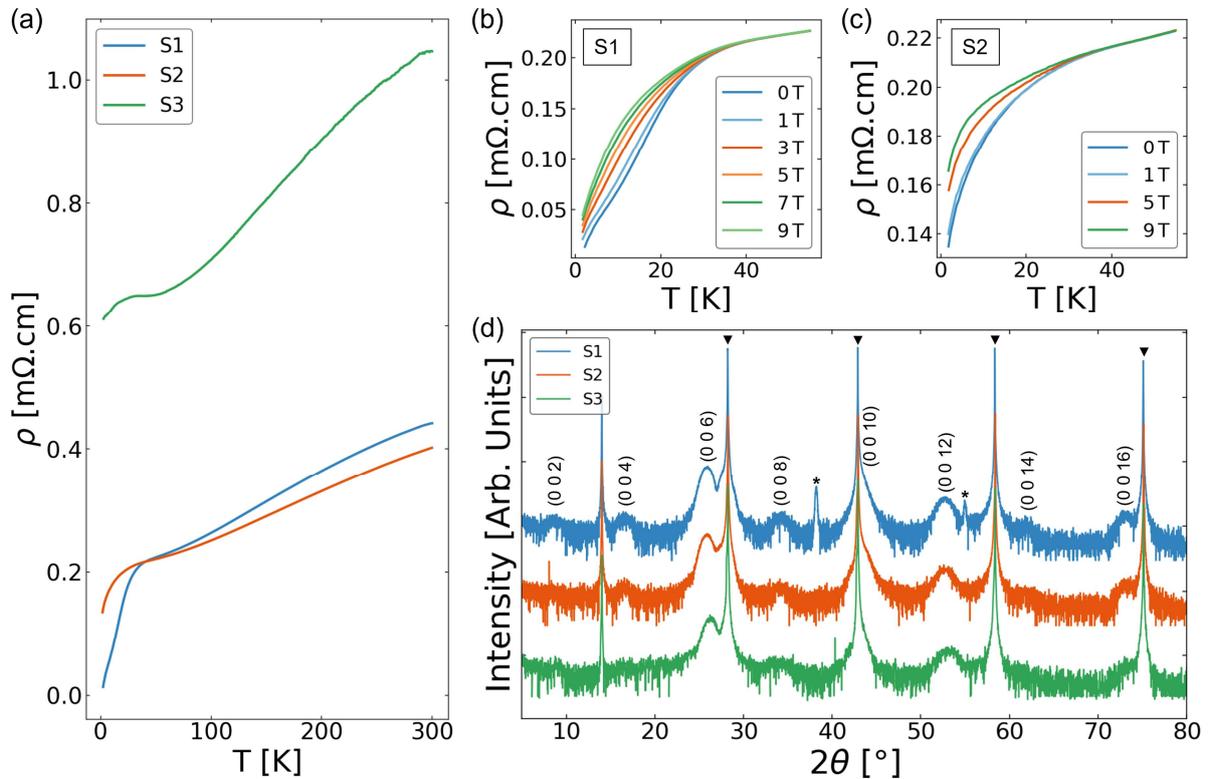

Figure 2: **a)** Temperature-dependent resistivity measurements of S1, S2 and S3 after $O_3$ annealing. **b),c)** Magnetic field dependence of the low temperature resistivity of (b) S1 and (c) S2. (d) X-ray diffraction patterns of S1, S2 and S3 after their respective $O_3$ annealing. The characteristic peaks of LNO327 are labelled as (00l). The ▼ and ∗ symbols correspond to the substrate and Au top electrodes respectively.

Table 1: Ozone annealing parameters for a set of equivalent LNO327 samples and related critical temperature calculated as describide in the text.

| Sample | temperature (°C) | time (min) | weight (%wt) | onset $T_C$ |
|---|---|---|---|---|
| S1 | 330 | 60 | 2.0 | 42 |
| S2 | 375 | 60 | 2.5 | 33 |
| S3 | 300 | 15 | 0.5 | – |

## 2.3 Atomic-Scale Structural Analysis via STEM-EELS

**Figure 3** shows the STEM–EELS characterization of sample S1. The HAADF-STEM image (Figure 3a) shows a *ca.* 6 nm thick $La_3Ni_2O_{7-\delta}$ layer visible in bright contrast. The nickelate region exhibits a well-developed LNO-2222 structure over most of the field of view, with a nascent defective region at the bottom of the image. To better study those different regions we extracted several numerical diffractograms from this HAADF-STEM image and reported them in Figures 3b,c,d. The first diffractogram (blue box), displays clearly defined Bragg spots, with intense reflections indexed as (006), (220), and (115), consistent with the *Amam* symmetry of the LNO-2222 structure. In agreement with the XRD pattern shown in Figure 1b, periodicities corresponding to (002), (004), ... up to (0016) are also visible in the diffractogram obtained from the STEM images. The HAADF-STEM image in Figure 3e was acquired from a well-developed

LNO-2222 region where EELS measurements were also performed. Since the La-$M$ and Ni-$L$ edges partially overlap, component unmixing was applied (please refer to the SI). Figure 3f shows the EELS spatial map associated to the spectral component of the Ni-$L$ signal, while Figure 3g corresponds to the raw intensity of the La-$M$ edge. This spectroscopic analysis reveals a clear 1–2–2–2–2–2–1 Ni-layer stacking sequence from the top interface toward the bottom interface with the SLAO substrate.

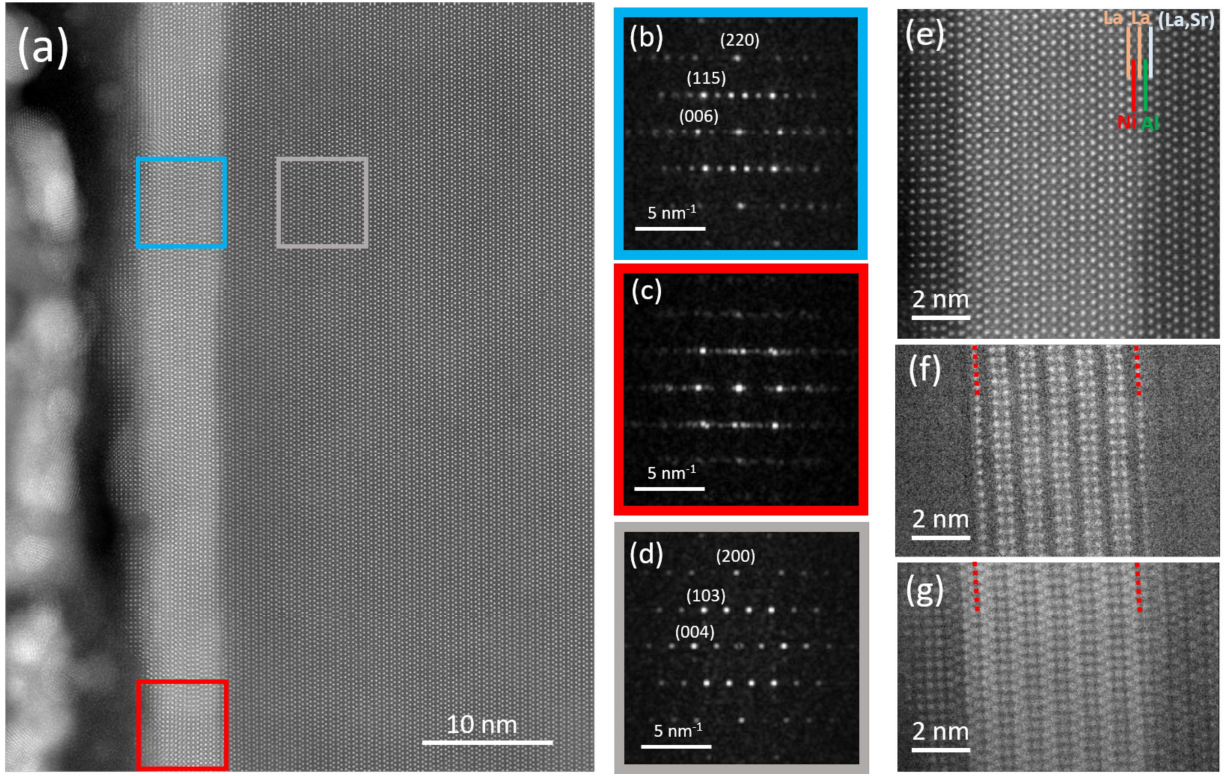

Figure 3: (a) HAADF-STEM image of the S1 sample showing a $La_3Ni_2O_7$ layer (bright contrast) on top of a $SrLaAlO_4$ substrate. (b-d) Numerical diffractograms extracted from the HAADF image on the $La_3Ni_2O_7$ layer, defective $La_3Ni_2O_7$ layer and the $SrLaAlO_4$ substrate. The boxes in (a) indicate the areas used to obtain the diffractograms. (e) HAADF image acquired in a well-developed 2222 region used for EELS. The plane sequence at the $La_3Ni_2O_7$ / $SrLaAlO_4$ interface is indicated. (f) Ni-$L$ edge components after EELS spectral unmixing. (g) Raw EELS La-$M$ edge intensity map. The red dotted lines indicate the first and last nickel plane positions.

Interestingly, the first Ni plane at the interface with the substrate belongs to a bilayer block and the EELS chemical map indicates that this Ni plane is surrounded by La-rich cation layers. Since the sample is grown onto SLAO, the second part of this BL structure is expected to consist of Al and a mixed (La,Sr) plane, as no Sr/La contrast modulation is observed by HAADF or EELS in the substrate. This results in an interfacial cation sequence of La-Ni-La-Al-(La,Sr), corresponding to an interface composed of a bilayer $La_{2.5}Sr_{0.5}NiAlO_7$, where a $Ni^{2.5+}$ oxidation state is preserved as in nominal $La_3Ni_2O_7$. At the top interface, Ni atoms adopt a $La_2NiO_4$-type structure, while the top SLAO layer exhibits numerous faults and some La enrichment, preventing a clear determination of the Ni valence. In the lower part of Figure 3a, the stacking sequence appears faulted with a ML-TL sequence, in connection with a dislocation near the interface. The corresponding diffractogram (red box) shows that while reflections corresponding to (220), (115), (006), and (0010) remain visible, the longest periodicities disappear in this defective region. The profiles obtained across the (00$l$) line in the diffractogram for the two regions are compared in the SI (please refer to the SI). The observed differences are similar to those seen in the XRD patterns of S1 and S3, clearly revealing that disorder and formation of stacking faults, such as LNO-1313 intergrowth, can explain the changes observed for the XRD of S3 after the ozone annealing. The diffractogram of the substrate (grey box) corresponds to the monolayer stacking sequence, as expected for the $n = 1$ RP structure of $SrLaAlO_4$. In those Ruddlesden–Popper structures, similar periodicities occur across different members of the series.

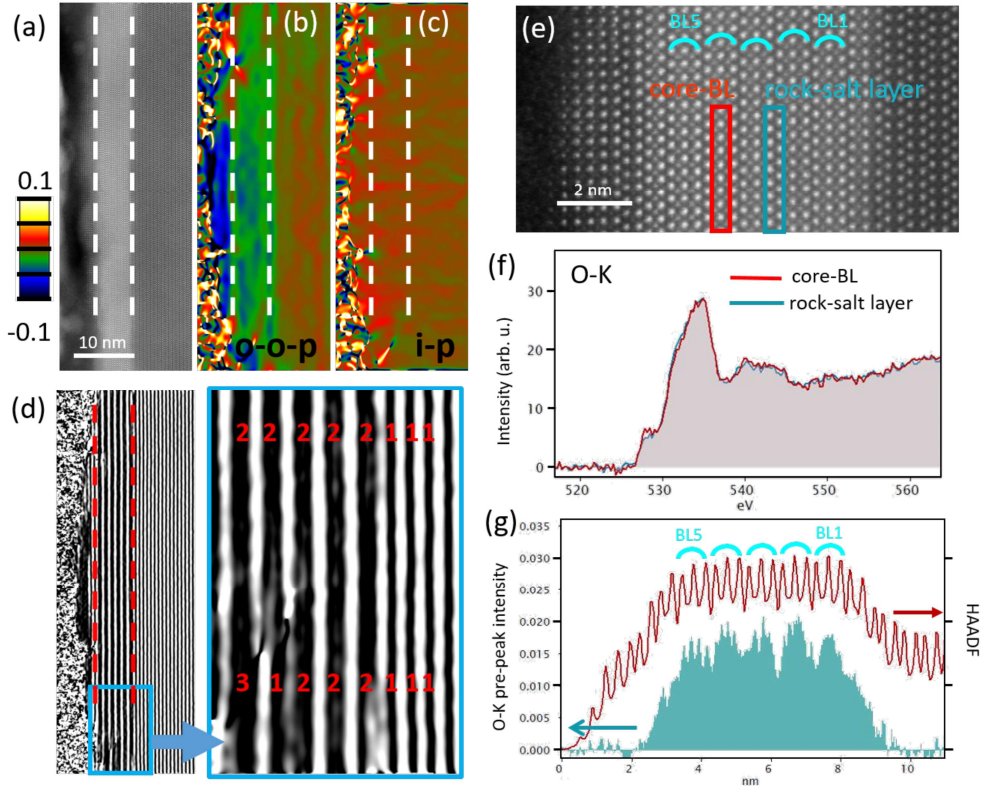

Figure 4: **(a–c)** HAADF-STEM image of sample S1 with out-of-plane (o-o-p) and in-plane (i-p) lattice expansion maps from GPA. No rotation or shear relative to the substrate was detected, the corresponding maps are therefore not displayed. Color scale for panels (b–c) ranges from −10% to +10%. **(d)** Ruddlesden–Popper stacking sequences from high-resolution GPA; white contrast indicates the phase shift associated with A-site cation doubling. **(e)** HAADF-STEM image of a defect-free zone showing five successive bilayers and a well-crystallized capping layer. **(f)** O–K EELS edges extracted from the rock-salt layer and the perovskite block, as indicated in (e). **(g)** O–K pre-peak intensity profile across the nickelate layers with the corresponding HAADF-STEM profile.

For example, the (0010) reflection of the $n=2$ structure (either 2222 or 1313 polytypes) corresponds to the (006) reflection of the $n=1$ structure, both associated with an interplanar spacing of approximately 0.20-0.21 nm, corresponding to the cation layer spacing along the $c$-axis. Similarly, the (115) reflection of the LNO-2222 phase corresponds to the (103) reflection of the $n=1$ structure. These correspondences allow determination of the unit cell parameters through geometrical phase analysis (GPA), using either the (006) and (200) reflections or the (103) and (−103) reflections of the substrate as references.

In sample S1, the LNO-2222 structure is largely defect-free over lateral distances of approximately 50 nm (*cf.* **Figure 4a**). The corresponding out-of-plane (o-o-p) and in-plane (i-p) lattice expansion maps (Figure 4b,c), confirm a well-preserved epitaxial i-p strain, with only a minor unit-cell expansion observed near the top of the film up to ∼ 0.4%. The o-o-p map further reveals relatively homogeneous lattice parameters across the LNO-2222 region, with values approximately 2% larger than that of the substrate. The resulting $c$-axis parameter ranges between 20.53 Å and 20.74 Å (for a SrLaAlO$_4$ reference unit cell of 12.63 Å), in good agreement with the values obtained from XRD within the error bars. Using a GPA aperture corresponding to ∼ 0.4 nm spatial resolution, phase discontinuities associated with RP stacking can be imaged. Figure 4d confirms the doubling of the RP stacking periodicity in the La$_3$Ni$_2$O$_7$ layer relative to the substrate. In regions where stacking faults are present (blue box in Figure 4d), GPA reveals LNO-1313 polytypes and locally the RP sequence becomes 1–3–1–2–2–2–2 instead of 1–2–2–2–2–2–2. Nevertheless, this analysis confirms that sample S1 exhibits a rather homogeneous structure, with pure LNO-2222 domains extending over lateral distances of ca. 50 nm and across the entire film thickness without any apparent heterogeneity. Possible electronic variations in these apparently homogeneous regions were probed by EELS measurements at the O–K edge (Figure 4e. As shown in Figure 4f, the O–K pre-edge displays subtle differences when comparing the rock-salt layers to the core of the bilayer, *i.e.*, the perovskite-type block. In the latter, the pre-peak exhibits a stronger intensity, consistent with enhanced Ni-O hybridization. Such differences have already been reported for $La_3Ni_2O_7$ single crystals [24]. Figure 4g shows the evolution of the pre-peak intensity, normalized to the main O–K edge, across five perfect bilayers in the film. The pre-peak intensity follows the expected atomic-scale modulation between rock-salt and perovskite layers, but remains relatively constant from one bilayer to another. In particular, no significant reduction of the pre-peak is observed for the LNO-2222 near the top of the film. This trend is often observed when the $SrLaAlO_4$ capping layer is well crystallized, as in this case. In contrast, when the capping layer is absent or poorly ordered, the upper part of the nickelate layer often exhibits a reduced O-K pre-edge intensity, likely associated with a decrease in Ni valence due to oxygen diffusion and loss (please refer to the SI). A similar effect has been reported as the origin of the conductivity gradient across the film thickness, especially for uncapped films ([25]).

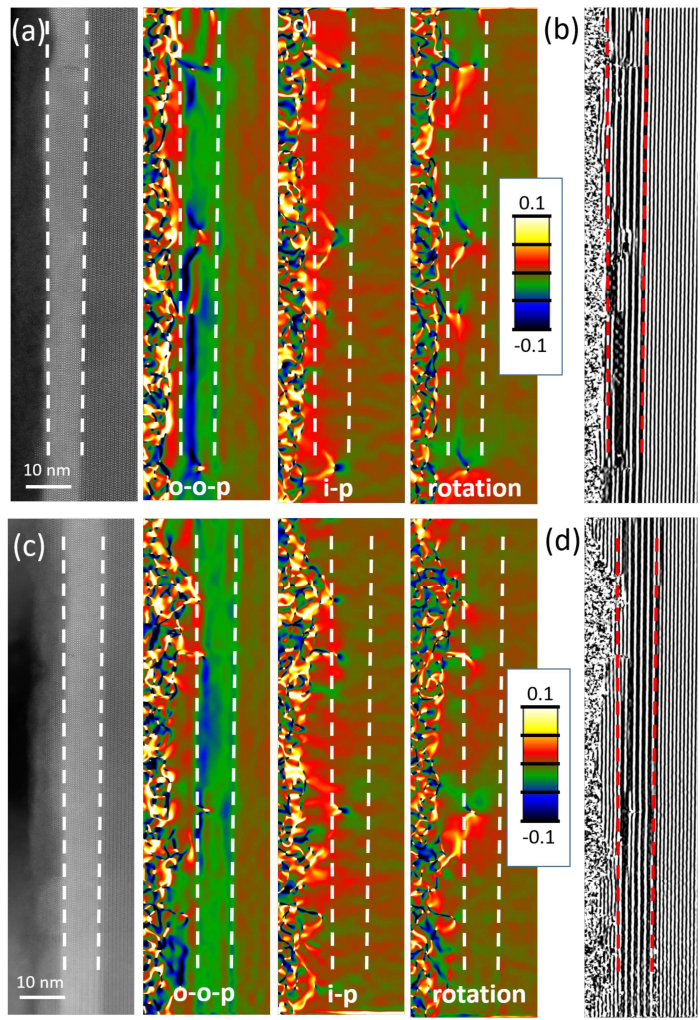

Figure 5: (a) HAADF-STEM image of sample S3 with out-of-plane (o-o-p), in-plane (i-p) lattice expansion maps and rotation map from GPA (b) Ruddlesden–Popper stacking sequence from high-resolution GPA of sample S3. (c,d) Similar HAADF-STEM and GPA analysis of the S2 sample. Color scale for panels (a,c) ranges from $-10\%$ to $+10\%$ for the lattice expansion maps and from $-10°$ to $+10°$ for the rotation map.

HAADF-STEM investigations were also carried out on samples S2 and S3, as shown in **Figure 5**, with GPA analyses performed in the same manner as for S1 to ensure consistent comparison. Sample S3 exhibits a significantly higher density of stacking faults, as illustrated in Figure 5a where the LNO-2222 stacking is strongly disrupted. Across the observed areas of the FIB lamellae, the LNO-2222 phase is mainly confined near the substrate interface and almost never extends through the entire layer thickness. Various faults are present, including the insertion of monolayer "1" sequences or the merging of "22" blocks into local "4"

RP-type phases. The LNO-1313 polytype is also frequently observed. These defects suppress the (002) and (004) reflections in the diffractograms, and the out-of-plane lattice parameter varies locally, consistent with the presence of 1313 domains, whose $c$-axis parameter is smaller than that of the 2222 structure ($cf.$ o-o-p lattice parameter imaging in Figure 5a). In addition, the i-p strain is partially relaxed, as indicated by the GPA analysis. This relaxation is associated with the numerous structural faults observed in Figure 5b. As further confirmed by the rotation map, these defects introduce structural heterogeneities, often starting in the first half of the nickelate layer and resulting in structural impacts even within the LNO-2222 regions. The STEM structural investigation of sample S2 is shown in Figure 5c,d. It also exhibits disorder and stacking faults, but to a much lesser extent than sample S3. Notably, the LNO-2222 polytype often extends across the entire film thickness. Defects are also visible in the GPA images, but they are primarily confined near the top interface of the LNO layers, resulting in a reduced impact on the film strain and leaving a substantial portion of the LNO-2222 structure largely strained to the substrate. Nevertheless, taking advantage of the heterogeneity still present in this film, an O-K edge EELS characterization of the different polytypes could be performed. **Figure 6** shows that the O-K pre-peak intensity is modulated depending on the polytype. The O-K pre-peak intensity reflects the degree of Ni–O hybridization and the Ni valence state, and increases with both the degree of hybridization and the Ni valence. This is particularly relevant in this system, where the ground state is characterized by a substantial charge-transfer from the O-2p bands towards the unoccupied Ni–3d states. For the bulk Ruddlesden–Popper series $La_{n+1}Ni_nO_{3n+1}$, the Ni valence is expected to be $Ni^{(3-1/n)+}$. Consistent with changes in valence and increased connectivity of the Ni–O network, the O-K pre-peak intensity strongly increases when measured from an area encompassing two neighboring monolayers (denoted "11") to a block corresponding to a unit cell of the $n=4$ RP structure (see Figure 6 and more details in the SI). The LNO-2222 and 1313 polytypes are expected to have nominally similar valence states, and the EELS measurements on "22" and "13" areas show rather similar O-K pre-edge intensities. Nevertheless, spectra extracted from the ML and TL of the "13" block exhibit clear differences: the ML shows a much weaker pre-peak shifted to higher energy. Although subtle, a shift towards higher energies is also observed for the O-K pre-peak onset of the "22" area ($ca.$ 0.5 eV), when compared to the "13" or "4" blocks. This may indicate a small band-gap opening in those "22" regions. While speculative, this could suggest that the "22" layer is more sensitive to ambient conditions than other polytypes, possibly due to its higher susceptibility to changes in oxygen stoichiometry. This interpretation is consistent with previously reported momentum-integrated energy distribution curves near the Fermi level, which reveal a conductivity gradient across the LNO-2222 film thickness, particularly in uncapped films, where some bilayers exhibit insulating-like behavior, while those near the substrate interface retain metallic character [25]. This is also consistent with our observation that the LNO-2222 structure is prone to be suppressed under $O_3$ treatment, as seen in sample S3, and supports the conclusion that the establishment of superconductivity is strongly dependent on the atmosphere and duration of the sample aging. These systematic variations of the EELS O-K edge with the local nickelate structure demonstrate that stacking faults and different amounts of polytypes directly affect the electronic structure. In this context, comparison of the three samples shows that only sample S1 exhibits not only structural but also apparent electronic homogeneity extending over large areas of several tens of nanometers.

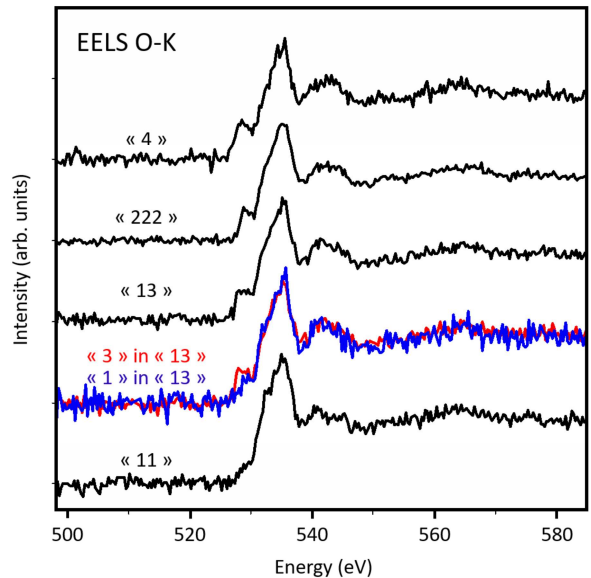

Figure 6: EELS O–K edge spectra measured at different locations of the thin film, corresponding to different stacking layers of the RP polytype. '1', '2', etc., correspond to monolayer, bilayer, and higher-order layers. The '11', '13', etc., indicate measurements encompassing respectively two neighboring monolayers or a neighboring monolayer and trilayer, and so on.

Finally, we emphasize that the structural differences observed among our samples can be rationalized by considering the $O_3$ oxidizing power under varying annealing conditions. For sample S2, annealed at 375°C, the relatively low oxidizing power of $O_3$ at this temperature necessitates prolonged annealing times to achieve full oxidation of the film. However, such extended exposure risks compromising the crystalline structure, a degradation that we indeed observe experimentally. Conversely, sample S3, annealed at a lower temperature (300°C), experiences a higher relative $O_3$ oxidizing power, which likewise induces modifications in its crystalline structure. This delicate interplay between annealing time and oxidation power underscores the challenge of simultaneously achieving optimal oxidation and structural integrity in those bylayer nickelate thin films. Our observations align with the recent work of Liu et al. [19], who constructed an experimental phase diagram mapping the $O_3$ oxidizing power as a function of the annealing temperature for $(La,Pr)_3Ni_2O_7$ thin films, further validating the intrinsic complexity of this balance.

## 3 Conclusion

Our study reveals a clear correlation between the structural polytype of $La_3Ni_2O_{7-\delta}$ thin films and their superconducting properties. We observe that robust superconductivity emerges in samples dominated by the LNO-2222 polytype, particularly when it extends over the largest lateral domains and throughout the full thickness of the layer, highlighting its critical role in establishing a coherent superconducting state. In contrast, samples with a higher volume fraction of alternative polytypes exhibit poorer transport properties. HAADF-STEM and EELS analyses of the O-K edge suggest that those polytypes may disrupt the structural and electronic coherence required for optimal superconductivity. The question of whether superconductivity can be achieved in the LNO-1313 phase, as demonstrated under pressure in the literature, remains unresolved. Although this polytype is present in large volume fractions in the non-superconducting samples and could therefore be associated with the absence of superconductivity, the excessively high defect density, independent of LNO-1313, most likely governs the observed behavior. The difficult to obtain phase-pure and defect-less LNO-1313 samples nevertheless suggests that targeting the LNO-2222 phase may be the most promising route to achieve superconductivity. This phase appears to be particularly sensitive, structurally and likely electronically, to ozone treatment and environmental conditions. Future efforts should thus focus on optimizing growth parameters to minimize defects and stabilize the 2222 phase, for example, through the use of a well-crystallized capping layer, as demonstrated in our work, thereby paving the way for reproducible and robust superconductivity in this material system.

## 4 Experimental Section

*Growth, structural and transport characterizations*:
The epitaxial growth of LNO327 thin films onto (001)-oriented $SrLaAlO_4$ (SLAO) substrates was done by pulsed laser deposition using a 248 nm KrF excimer laser and utilizing stoichiometric ceramic targets from Toshima Manufacturing CO LTD. The SLAO substrates, from CODEX international, were sonicated in actone and isopropanol for 10 minutes each. The growth was performed using a temperature of 685°C, an oxygen partial pressure of 0.1 mbar and a fluence of 2.97 J.cm$^{-2}$. After the deposition of the LNO327 layer, the samples were capped by a 6 nm SLAO layer under the same temperature/pressure conditions.

The surface morphology of the samples was acquired using a Park XE7 (Park System) atomic force microscope (AFM) in true non-contact mode.

The XRD data were collected using a Rigaku SmartLab diffractometer equipped with a rotating anode Cu-K$\alpha$ radiation source ($\lambda$=0.154056 nm). Transport measurements were performed by using a Dynacool system (Quantum Design) in a van der Pauw method on 3×3 mm$^2$ samples by applying current amplitudes of 10 $\mu$A. For some of the samples also square Au top electrodes were evaporated.

*Ozone annealing*:

The post-growth ozone annealings were performed in an home-made tubular furnace connected to an ozone generator BMT 803 BT from BMT Messtechnik GmbH at different temperatures as described in Table 1. The O$_2$ flow was kept constant and equal to 1600sccm or 800sccm and the O$_3$ concentration was controlled by a potentiometer. Prior the gas insertion, the tube was put under a vacuum of about 0.2 mbar in order to ensure a pure oxidative environment. The heating up of each annealing process was performed with a very small ramp rate of circa 5°/min, contrarily to what done in literature [19]. Finally, the samples were removed from the furnace when the temperature was below 200°C.

*High resolution STEM-EELS*:

Cross-sectional transmission electron microscopy (TEM) lamellae were prepared using a focused ion beam (FIB) technique (C2N, University of Paris-Saclay, France). Before FIB lamella preparation, approximately 20 nm of amorphous carbon was deposited on top for protection. The samples were observed in the days following FIB preparation. Several lamella thickness windows were prepared during the FIB process, with typical thicknesses ranging from 30 to 100 nm, as estimated by EELS.

HAADF imaging was carried out using a NION UltraSTEM 200 C3/C5-corrected scanning transmission electron microscope (STEM). Experiments were performed at 100 keV and 200 keV with a probe current of approximately 10–30 pA and convergence semi-angles of 30 mrad. No significant differences were observed between 100 keV and 200 keV experiments, particularly in EELS, indicating that the conditions were sufficiently gentle to avoid oxygen sputtering. For the EELS, a MerlinEM detector (Quantum Detectors Ltd) in a 4 × 1 configuration (1024 × 256) was installed on a Gatan ENFINA spectrometer mounted on the microscope [26], enabling electron-counting detection efficiency.

Geometrical phase analysis (GPA) [27] was performed with a spatial resolution of 1.5 nm for lattice expansion and rotation maps, and 0.4 nm for imaging the Ruddlesden–Popper stacking sequence. The O–K intensity profiles were obtained by integrating the O–K pre-edge from 526.5 eV to 530.5 eV and normalizing it to the intensity integrated from 534 eV to 617 eV. For a given polytype, we observe that the O–K edge is sensitive to both the presence of the capping layer and the local thickness. Thinner regions (ca. 30 nm) show reduced O–K pre-peak intensity; therefore, for consistency, all O–K edge analyses discussed in the manuscript correspond to similar thicknesses (approximately 60 nm).

**Acknowledgements**

This work was funded by the French National Research Agency (ANR) through the project ORBIFUN (ANR-23-ERCS-0003). MF and NV acknowledge financial support from the ANR ACCURATE (ANR-23-CE08-0017). AG acknowledges financial support from the ANR ImagingQM (ANR-23-CE42-0027). This work was also done as part of the Interdisciplinary Thematic Institute QMat, ITI 2021 2028 program of the University of Strasbourg, CNRS and Inserm, and supported by IdEx Unistra (ANR 10 IDEX 0002), and by SFRI STRAT'US project (ANR 20 SFRI 0012) and EUR QMAT ANR-17-EURE-0024 under the framework of the French Investments for the Future Program. It was also supported by France 2030 government investment plan managed by the French National Research Agency under grant reference PEPR SPIN – [SPINMAT] ANR-22-EXSP-0007. The authors thank the PLD, XRD, MEB-CRO and MagTransCS platforms of the IPCMS.

# Supporting Information (SI) for

# Decoding Superconductivity in La$_3$Ni$_2$O$_{7-\delta}$ Thin Films via Ozone-Driven Structure and Oxidation Tuning

*Mathieu Flavenot, Hoshang Sahib, Jerome Robert, Marc Lenertz, Gilles Versini, Laurent Schlur, Alexandre Gloter\*, Nathalie Viart* and *Daniele Preziosi\**

M. Flavenot, H. Sahib⊥, J. Robert, M. Lenertz, G. Versini, L. Schlur, N. Viart and D. Preziosi
Université de Strasbourg, CNRS, IPCMS UMR 7504, F-67034 Strasbourg, France
Email Address: daniele.preziosi@ipcms.unistra.fr
A. Gloter
Laboratoire de Physique des Solides, CNRS, Université Paris-Saclay, 91405 Orsay, France
Email Address: alexandre.gloter@universite-paris-saclay.fr

⊥ Present Address : Department of Physics, College of Science, University of Halabja, Halabja, Iraq

### Transport properties of La$_3$Ni$_2$O$_{7-\delta}$ thin films onto SLAO prior O3 annealing

Figure S1 displays the temperature-dependent resistivity curve of a SLAO-capped LNO327 thin film onto a SLAO substrate. The sample shows a fully insulating behaviour in the entire temperature range accessible. According to Zhang *et al.* this behaviour is expected for an oxygen off-stoichiometry $\delta \geq 0.08$ Ref. [1].

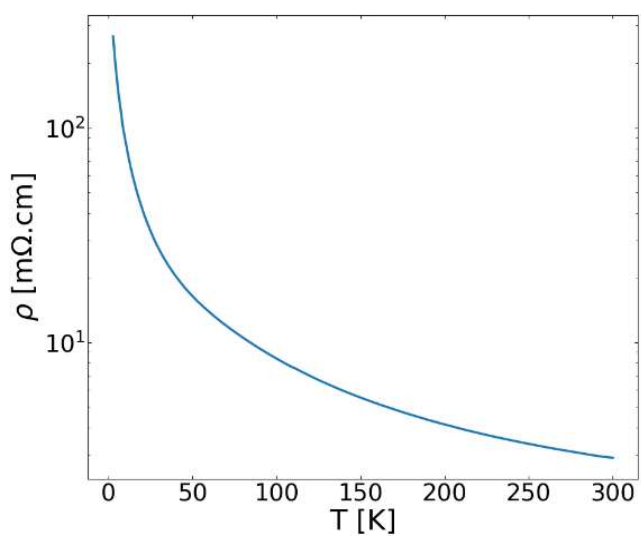

**Figure S1** Temperature dependent resistivity of La$_3$Ni$_2$O$_{7-\delta}$ thin film onto SLAO after the growth.

### Resistance inhomogeneity of S3

The temperature dependence of each sample was measured by using the Van der Pauw configuration. The measurements were performed by injecting 10µA of current in two opposite direction of the sample as sketched in Figure S2a,b. This allows to study the homogeneity of the transport properties. Figure S2c show that indeed sample S3 is highly inhomogeneous. Whereas measurements along the two directions show a fully metallic behaviour, the overall resistivity values show almost a factor 3 of difference. Moreover, their behaviour at low temperature differ completely: in one case we observe an upturn followed by a downturn reminiscent of a superconducting transition, in the other direction we observe only a minor upturn.

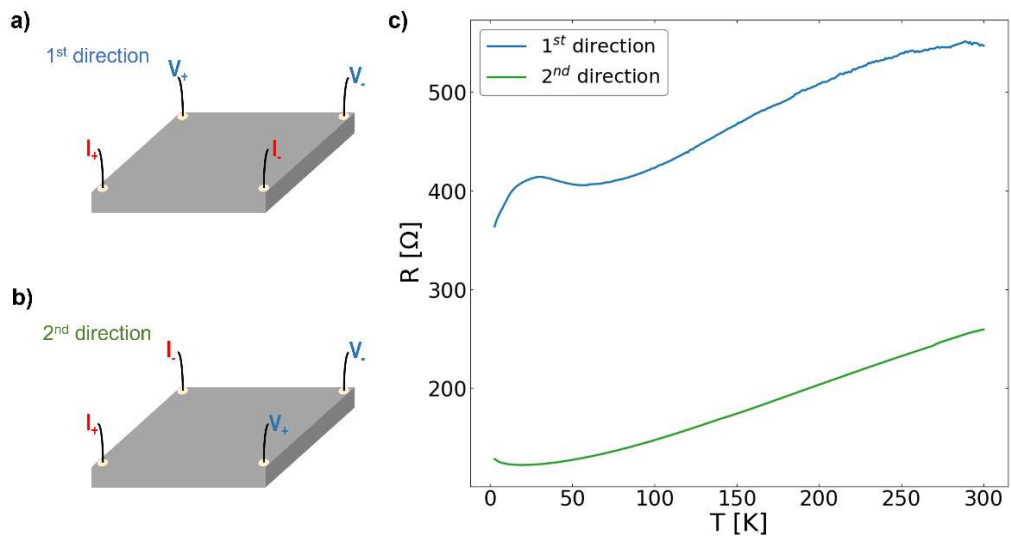

**Figure S2** a) Temperature dependent resistance of S3 in the two Van der Pauw configurations sketched on b) and c).

*Determination of the onset $T_c$ values*

The onset $T_c$ value is calculated by using the parallel-resistor model, which has previously been employed to describe the normal-state behavior of overdoped cuprates as well as bilayer nickelates thin films[2–5]. The resulting fits are shown in Figure S3 for samples S1 and S2, where the calculated curves closely reproduce the experimental resistivity data within the normal-state temperature range. The deviation between the experimental data and the fitted normal-state behavior at low temperatures is then used to determine the onset of the superconducting transition. No fitting was performed for sample S3 due to its degraded transport properties in the low-temperature region.

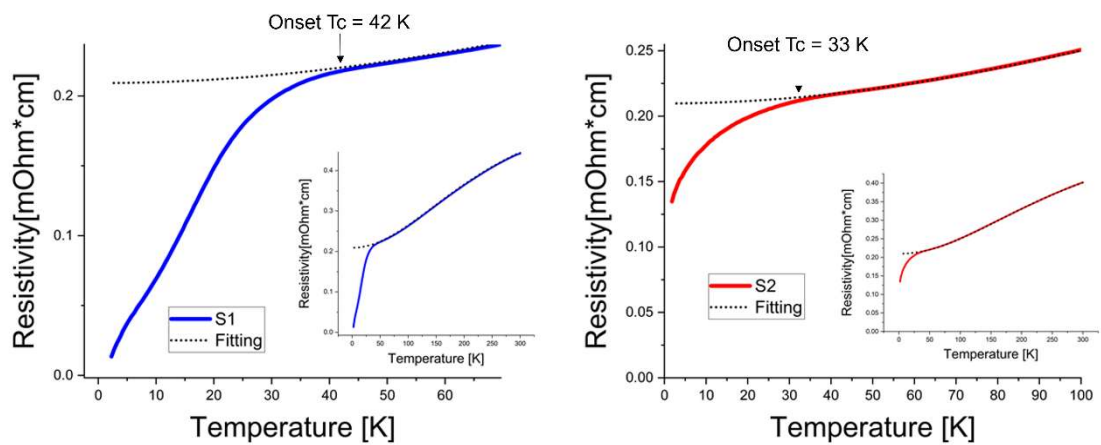

**Figure S3** Transport data for sample (left) S1 and (right) S2 superimposed to the fitting resulting from applying a parallel-resistor model to the normal-state resistivity.

*Evolution of the $T_c$ with a perpendicular magnetic field*

According to the linearized Ginzburg-Landau form on confined superconductor the critical field can be expressed as a function of the temperature as:

$$H_{c,\perp}(T) = \frac{\phi_0}{2\pi\xi_{ab}^2(0)}\left(1 - \frac{T}{T_c}\right)$$

$$H_{c,\parallel}(T) = \frac{\sqrt{12}\phi_0}{2\pi\xi_{ab}(0)d}\left(1 - \frac{T}{T_c}\right)^{\frac{1}{2}}$$

Where $H_{c,\perp}$ and $H_{c,\parallel}$ are the upper critical field in the perpendicular and parallel configuration respectively, $\phi_0$ is the flux quantum, $\xi_{ab}(0)$ the zero-temperature Ginzburg-Landau coherence and $d$ the superconducting thickness [need a ref]. Figure S4 presents the evolution of the critical field with the temperature in the perpendicular configuration for the samples S1 and S2. The critical field was extracted from the $T_c^{90\%}$, corresponding to the $T_c$ at 90% of the resistivity at the onset $T_c$ determined as shown in Figure S3. As expected by the Ginzburg-Landau theory, it can be modelled with a straight line. Moreover, the extracted upper critical of 87T and 25T for S1 and S2 respectively, points out to a more robust superconductivity for S1 as compared to S2.

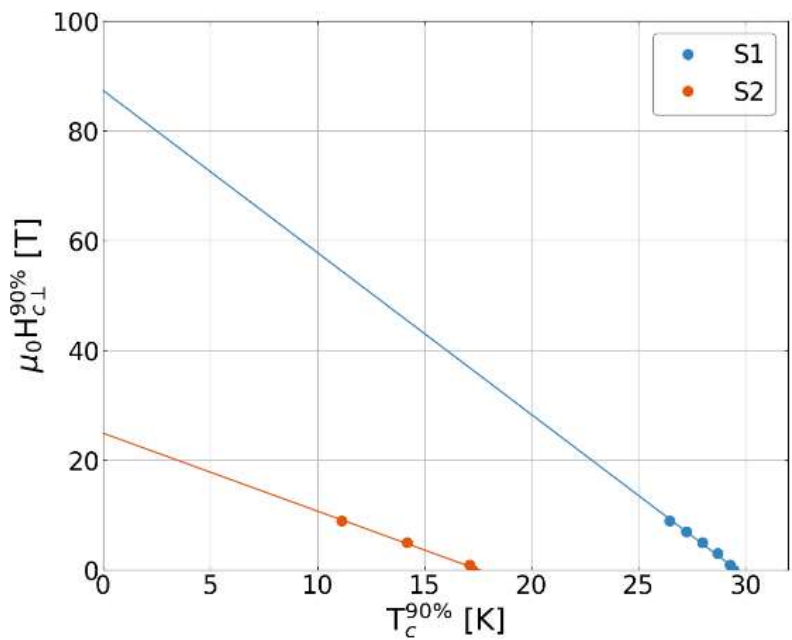

**Figure S4** Critical fields of S1 and S2 based on the $T_c$ at 90% of the normal state resistivity ($T_c^{90\%}$). The magnetic field was applied perpendicularly to the sample surface. The solid lines correspond to the linear fit.

*Hall effect of superconducting LNO327 thin films*

Figure S5 presents the Hall effect measurement at different temperature of a superconducting LNO327 thin film onto SLAO. The sample displays an onset $T_c$ around 38K as determined by the fitting of the normal-state transport (see the corresponding section). A positive Hall coefficient ($R_H$) is obtained throughout the measurement range (figure S5.b) corresponding to a hole behaviour, which is in agreement with the literature on undoped LNO327 thin film[6,7].

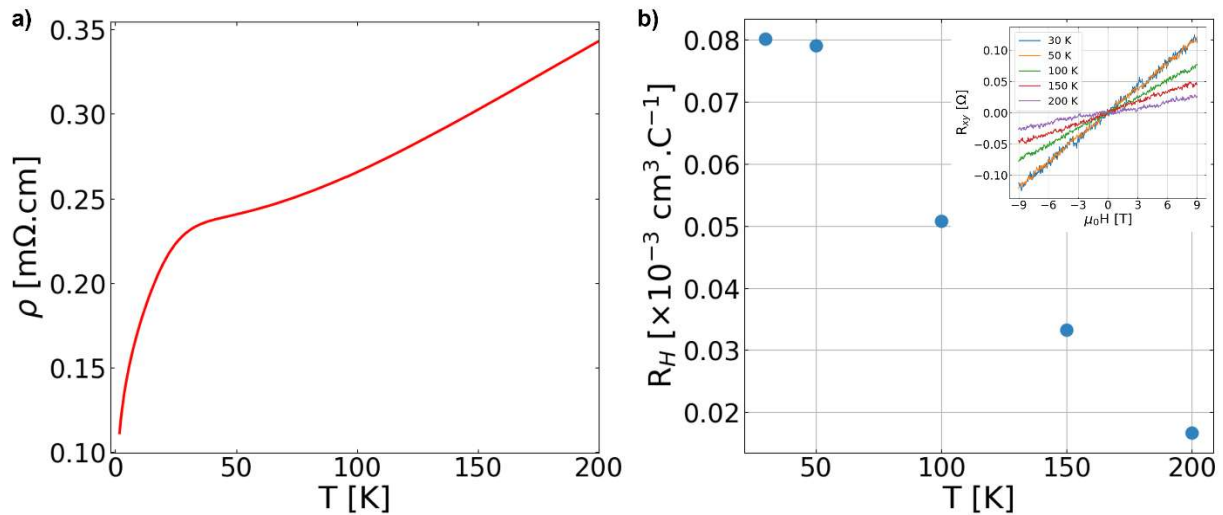

**Figure S5** a) Temperature-dependent resistivity measurement of a LNO327 thin film onto SLAO after ozone annealing. b) Temperature-dependent Hall coefficient of the sample presented in (a). The inset presents the Hall resistance as a function of the magnetic field.

*Ni elemental mapping using STEM–EELS*

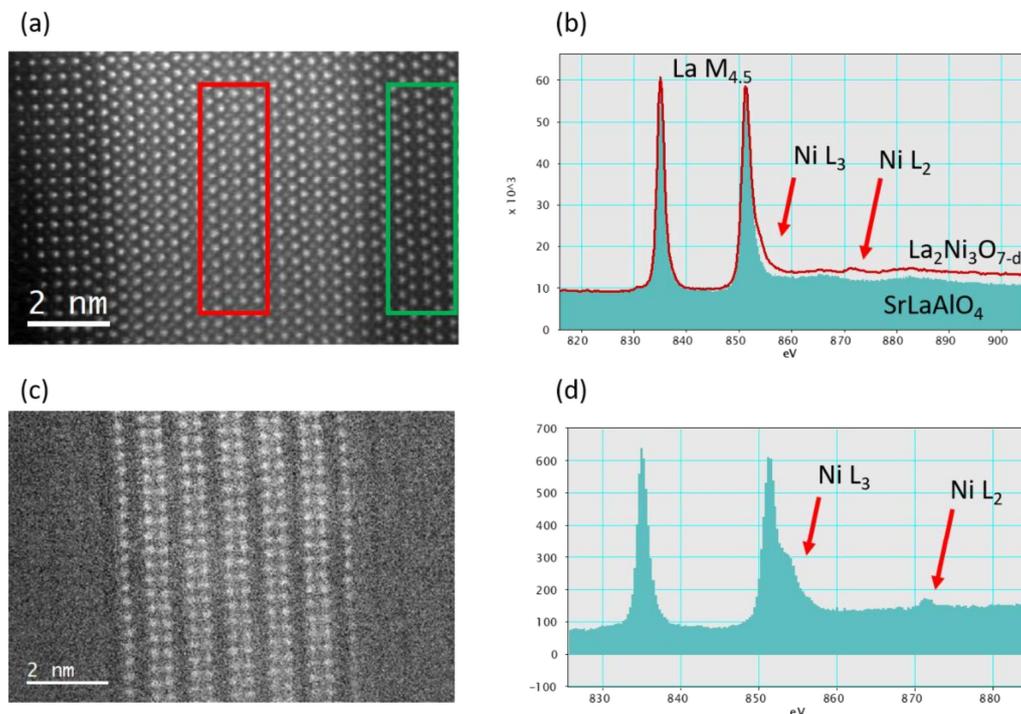

**Figure S6** a) HAADF-STEM image of sample S1, showing the same area as in Figure 3(e–g) in the main manuscript. b) Raw EELS data obtained by integrating the spectrum

**image data cube over the boxed areas in (a). c,d) Spatial distribution (c) of the spectral component shown in (d).**

Figure S6a presents the HAADF-STEM image of the sample S1. This image was acquired simultaneously with the EELS spectrum imaging. The raw EELS data on figure S6b was obtained by integrating the spectrum image from red and green rectangular areas in (a). These regions correspond to the inner part of the nickelate layer (LNO-2222) and to the substrate (SLAO). The spectra were normalized (background and intensity) to the signal before the La $M_5$-edge and to the maximum intensity of the La $M_4$-edge. The Ni $L_{2,3}$-edges are clearly visible atop the much more intense La edges. The data cube contains 363 × 238 spatial pixels × 1024 energy channels, corresponding to a 30 pm spatial step and ~0.3 eV per channel. The dwell time was 5 ms, and the incident beam current was ~30 pA. In Figure S6c,d are shown the spatial distribution (c) and its associated spectral component (d). This component was obtained after multivariate statistical analysis and component unmixing using the Varimax technique[6]. Here, the component, extracted from a decomposition into five spectral components, clearly corresponds to the Ni cation planes. Consistent results are obtained for decompositions using between 2 and 20 components, indicating that the Ni-plane mapping is largely independent of the number of components used. The spectral unmixing appears robust due to the high signal-to-noise ratio provided by the EELS detection system (Medipix 3). As a consistency check, similar Ni maps can also be obtained either by PCA denoising followed by edge subtraction using the "Egerton" method on the Ni $L_2$-edge, or by spectral fitting of the La $M_5$ and Ni $L_3$-edges.

***Attenuation of specific Miller reflections caused by stacking faults within the LNO327 layer***

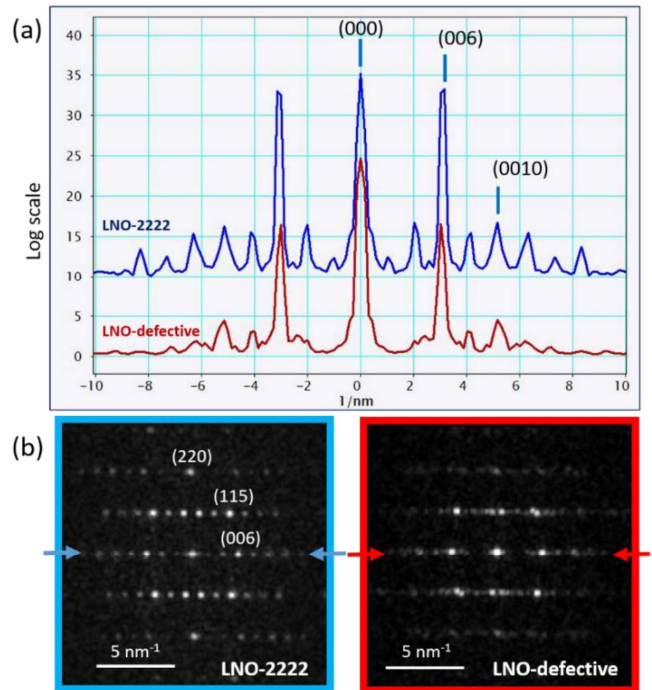

**Figure S7**  Profiles along the (0 0 l) Bragg line (a) from the diffractograms of a LNO-2222 (b) and a LNO-defective stackings (c). The arrows in (b,c) indicate the profile line directions.

Figure S7a shows the intensity profile of an ideal LNO-2222 and a LNO-defective diffractograms as shown in Fig. S7b. The diffractograms in (b) are the same as those described in Figure 3a,d of the main text. They correspond to the perfect LNO-2222 structure characterizing the S1 sample and to a region containing a nascent LNO-1313 defect embedded within neighbouring LNO-2222 layers. The arrows in (b) indicate the profile line directions. The profile in (a) shows that the (002), (004), (008), and planes above (0012) exhibit very faint intensity when LNO polytype defects begin to occur within the layer. The diffractograms were obtained from areas of approximately 10 × 10 nm in the HAADF-STEM images. For comparison, readers can refer to the bulk XRD in Figure 2d and to the evolution from films S1 to S3, where the presence of polytypes becomes more pronounced, accompanied by a similar disappearance of certain planes.

*Role of the capping layer on the EELS O K-edge and the Ni valence*

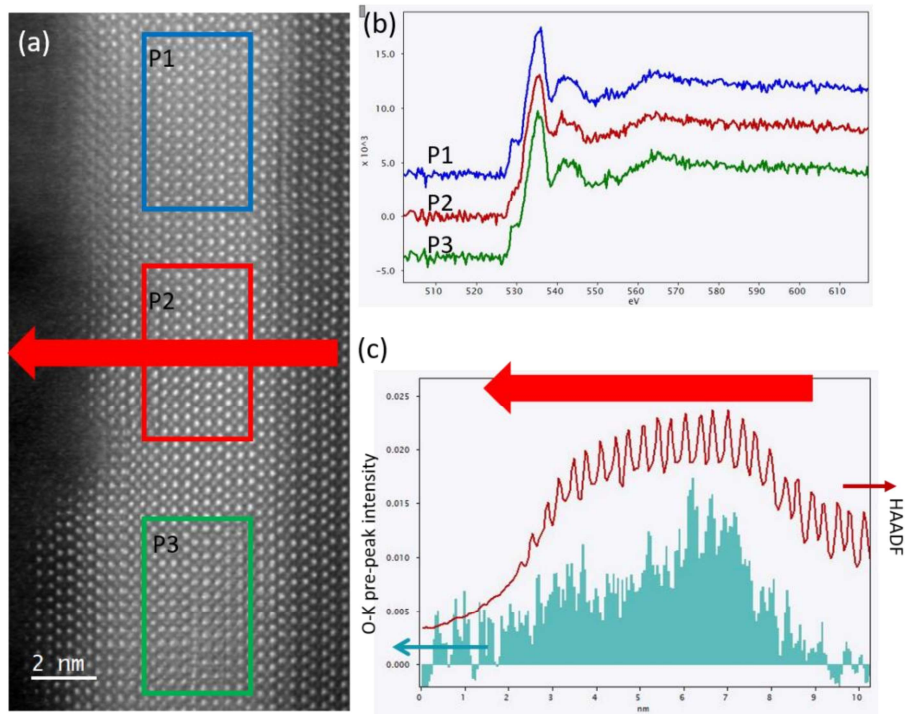

**Figure S8**   a) HAADF-STEM image of sample S1 presented in the main manuscript. b) Raw EELS data obtained by integrating the spectrum image data cube over the boxed areas in (a). c) Evolution of the O-K pre-peak along the sample thickness as represented by the red arrow on (a).

Figure S8a presents the HAADF-STEM image of the sample S1. This image was acquired simultaneously with the EELS spectrum imaging where the O K-edge is studied on three different areas represented by rectangular on (a). The EELS O-K signal in the red box (Position 2, P2) is less intense; however, the structure is equivalent to that in Position 1, *i.e.,* bilayer (BL)-type. Position 3, which includes a ML–TL ("13") stacking fault and a BL ("2"), exhibits a more intense O-K pre-edge than P2. In fact, the P2 area is affected by a poorly ordered or absent capping layer. When the capping layer is absent or defective, it is systematically observed that the O-K pre-edge is weaker, likely associated with oxygen loss and a reduction of the nickelate. This reduction is stronger near the top interface/surface of the nickelate. This is illustrated in Figure S8c, where the O-K pre-peak intensity is compared along with the HAADF-STEM image. In the upper half of the nickelate layer, the pre-peak is strongly reduced, indicating a reduction of the nickelate valence state.

*O–K EELS for different nickelate stacking*

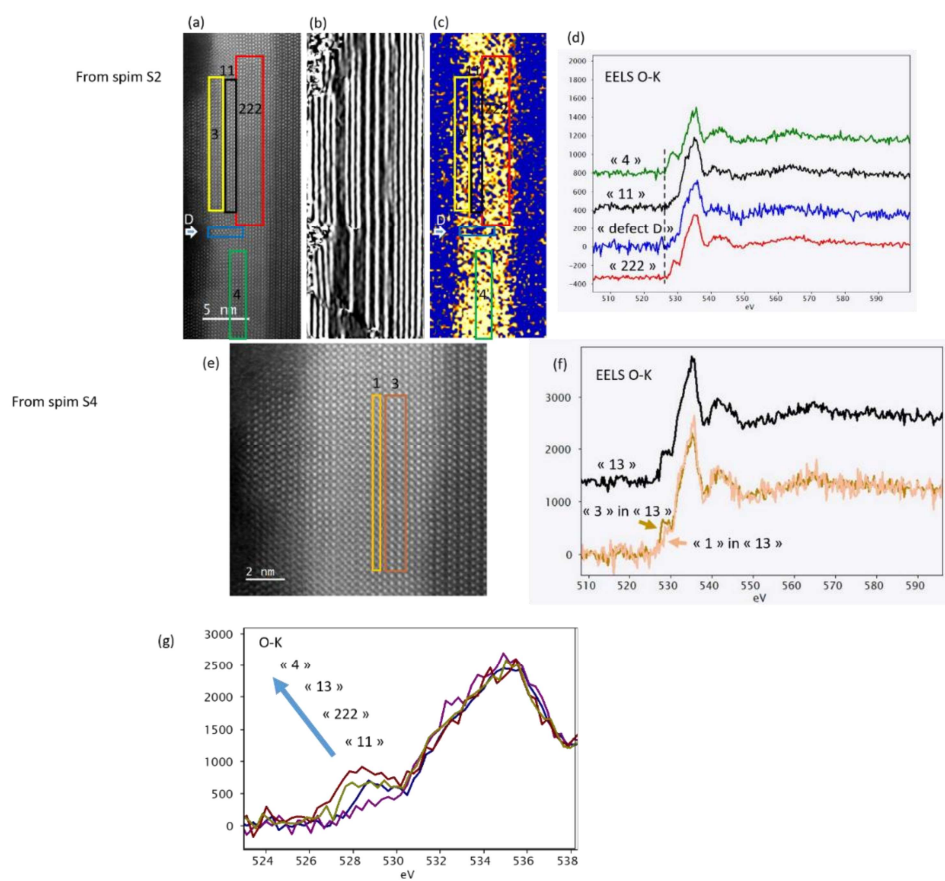

**Figure S9** a) HAADF-STEM image of sample S2 presented in the main manuscript. b) Ruddlesden–Popper stacking sequence obtained from high-resolution GPA analysis of the image in (a). c) Map of the O–K pre-peak intensity. d) O–K EELS spectra on the boxed area in (a,c). e) HAADF-STEM image of sample S2. f) EELS O–K spectra of the ML and TL extracted from the boxed area in (e). g) EELS O–K spectra, zoomed on the pre-peak region for the "11", "22", "13" and "4" areas.

Figure S9 a) shows the HAADF-STEM image of the sample S2 presented in the main manuscript. This image was acquired simultaneously with the EELS spectrum imaging on the O K-edge. Several boxes indicate typical stacking faults observed in this primarily LNO-2222 layer (red box). In particular, a TL–ML–ML sequence is highlighted (yellow-black boxes). A unit cell of the n=4 polytype is also observed (green box). It is also noteworthy that a defect composed of a rock-salt layer rotated by 90° separates different polytype regions (D and blue box). Above this defect, the capping layer is less ordered and thinner, indicating that the defect was already present prior to the deposition of the capping layer. On (b), is presented a Ruddlesden–Popper stacking sequence obtained from high-resolution GPA analysis of the image in (a). Figure S9 c) displays a map of the O–K pre-peak intensity. The signal-to-noise ratio is sufficient to resolve the O–K pre-peak intensity with sub-nanometer resolution. Bright yellow regions indicate higher O–K pre-edge intensity. The image shows that the area corresponding to the ML–ML region (black box) exhibits significantly weaker intensity. While closer to the top interface, the TL in the yellow box shows a stronger O–K pre-edge. The region corresponding to the n=4 Ruddlesden–Popper stacking sequence exhibits a further enhanced O–K pre-edge. The defective area D shows nearly no pre-edge intensity, consistent with a rock-salt layer and local degradation of the capping layer. Their corresponding O-K EELS spectras are shown on (d). It is noteworthy that the pre-edge onset of the LNO-2222 phase is slightly shifted toward higher energy (ca. 0.5 eV) compared to the n=4 polytype. The HAADF-STEM image of sample S2 presented on (e) was acquired simultaneously with the EELS spectrum imaging, the results of which are discussed in (f). A ML–TL structure is present in the first part of the layer, in regions where the capping layer remains intact, allowing, for instance, the O–K fine structure to be directly compared with the BL structure. The EELS signal integrated over the entire ML–TL region is also shown. Interestingly, the O–K pre-edge in the ML is much weaker and shifted to higher energy with respect to the TL. When compared to the bilayer (not shown here for simplicity), the ML–TL O–K onset appears at lower energy. This suggests that the LNO-2222 regions in the film exhibit an O–K onset slightly shifted to higher energy compared to the ML-TL and the n=4 polytypes. On (g) is shown the EELS O-K spectra zoomed on the pre-peak region for the "11", "22", "13", and "4" areas. As described in (d–f), the LNO-2222 pre-peak onset is located at lower energy compared to the "13" or "4" areas. Since the O–K onset position is an indirect fingerprint of the conduction band position or Fermi level in metallic oxides, this may indicate that the LNO-2222 phase tends toward a more semiconducting behaviour while other TL and n=4 will preserve the metallic behaviour. Although subtle, this observation suggests that the LNO-2222 polytype may be more sensitive than other polytypes and may more readily revert to a semiconducting state over time, during FIB preparation, partial strain relaxation, etc., whereas other polytypes may retain their metallic character more robustly. As mentioned in the main text, this interpretation is consistent with previously reported momentum-integrated energy distribution curves near the Fermi level, which reveal a conductivity gradient across the LNO-2222 film thickness, particularly in uncapped films, where some bilayers exhibit insulating-like behavior, while those near the substrate interface retain metallic character[2].